\documentclass{article}
\usepackage[latin1]{inputenc}
\usepackage{epsfig}
\usepackage{graphics}
\usepackage{amsmath,amssymb}
\usepackage{lscape}
\usepackage{citesort}

\date{}
\onecolumn

\begin{document}

\title{Ground state cooling, quantum state engineering and
 study of decoherence of ions in Paul traps}

\author{F. Schmidt-Kaler, Ch. Roos, H. C. N\"agerl*, H. Rohde, S. Gulde, \\
A. Mundt, M. Lederbauer, G. Thalhammer, Th. Zeiger, P. Barton,  \\
L. Hornekaer**, G. Reymond,  D. Leibfried, J. Eschner, and R. Blatt \\
\small Institut f{\"u}r Experimentalphysik, Universit{\"a}t Innsbruck,\\
\small Technikerstra{\ss}e 25, A-6020 Innsbruck, Austria,\\
\small *Norman Bridge Laboratory of Physics, Caltech, Pasadena, USA\\
\small **Aarhus University, Denmark}

\date{\today}
\maketitle

\begin{abstract}
We investigate single ions of $^{40}Ca^+$  in Paul traps for
quantum information processing. Superpositions of the S$_{1/2}$
electronic ground state and the metastable D$_{5/2}$ state are
used to implement a qubit. Laser light on the S$_{1/2}
\leftrightarrow$ D$_{5/2}$ transition is used for the manipulation
of the ion's quantum state. We apply sideband cooling to the ion
and reach the ground state of vibration with up to $99.9\%$
probability. Starting from this Fock state $|n=0\rangle$, we
demonstrate coherent quantum state manipulation. A large number of
Rabi oscillations and a ms-coherence time is observed. Motional
heating is measured to be as low as one vibrational quantum in
190~ms. We also report on ground state cooling of two ions.
\end{abstract}


\section{Quantum information processing with trapped ions}
Trapped and laser cooled ions in Paul  traps are used for an
implementation of quantum information processing. Internal
electronic states of individual ions serve to hold the quantum
information (qubits) and an excitation of  common vibrational
modes provides the coupling between qubits, which is necessary for
quantum logic operations between qubits, more specifically, for
the realization of gate operations between two ions. The
Cirac-Zoller proposal \cite{CiracZoller} requires that initially
the ions are cooled to the ground state of motion and that the
whole system can be coherently manipulated and controlled. The
time scale of decoherence and coupling to the environment is
required to be much smaller than the time scale of coherent
manipulation.

The paper is organized as follows: In section~II, we outline the
techniques which are used for storing and detecting single ions in
Paul traps. Section~III decribes  high resolution laser
spectroscopy on the S$_{1/2} \leftrightarrow$ D$_{5/2}$
transition, which is the basis of coherent manipulation of the
internal electronic and external vibrational quantum state of a
trapped ion. Ground state cooling for one and two ions is reported
in section~IV. For a single ion, we reached 99.9~$\%$ of motional
ground state occupation. Most recently, to scale up our results
from one ion towards strings of ions, we have also shown ground
state cooling for two ions. After a single ion is has been cooled,
it is prepared for coherent manipulation (section~V) of the ion's
vibrational state, including the generation of Fock states. We
measured the coherence time of this process as well as the time
scale of motional heating. Our experimental finding of 1~ms
coherence time, and a heating rate of only one phonon in 190~ms,
shows good conditions for quantum information processing. Finally,
in section~VI, we investigate the maximum speed of gate operations
and find that with our system approximately 30 - 50 gate
operations are possible within the coherence time.


\section{Paul traps for single Calcium ions}
Ions are stored in Paul traps \cite{Paul} under ultra high vacuum
($10^{-10..11}$~mbar) conditions. In one of the experiments we use
a spherical quadrupole Paul trap with a 1.4 mm ring diameter
(Fig.~1.a) \cite{Roos99}, where we observe motional frequencies
($\omega_x$, $\omega_y$, $\omega_z$)/(2$\pi$) of up to (2.16,
2.07, 4.51) MHz along the respective trap axes. The axial
direction of the trap is denoted by $z$, the degeneracy of radial
directions $x$ and $y$ is lifted by small asymmetries of the
setup. Strings of ions are held in a linear trap (Fig.~1.b)
\cite{Naegerl,NaegerlA}, where we typically reach motional
frequencies of ($\omega_x$, $\omega_y$, $\omega_z$)/(2$\pi$)=
(2.0, 2.0, 0.7) MHz, e.g. three ions crystallize at an axial
distance of 7$\mu$m.

\begin{figure}
\begin{minipage}{0.69\linewidth}
\begin{center}
\epsfxsize = 0.98 \linewidth
\epsfbox{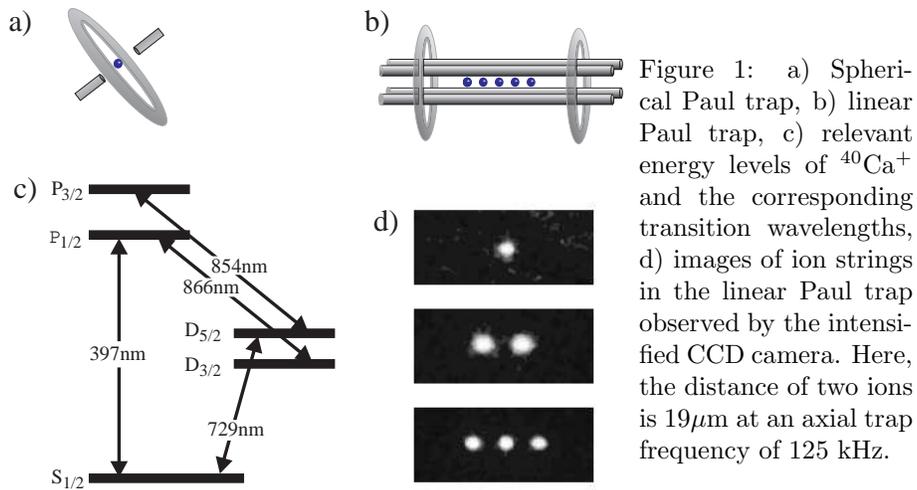}
\end{center}
\end{minipage}
\begin{minipage}{0.3\linewidth}
\caption{a) Spherical Paul trap, b) linear Paul trap, c) relevant
energy levels of $^{40}$Ca$^+$ and the corresponding transition
wavelengths, d) images of ion strings in the linear Paul trap
observed by the intensified CCD camera. Here, the distance of two
ions is 19$\mu$m at an axial trap frequency of 125~kHz.}
\end{minipage}
\end{figure}

$^{40}$Ca$^+$ ions have a single valence electron and no hyperfine
structure. All relevant transitions are accessible by solid state
or diode lasers (see Fig.~1.c) \cite{Roos99}. In our experiment,
we apply Doppler cooling to the ion and detect the internal state
on the S$_{1/2}$ to P$_{1/2}$ transition at 397~nm, excited with a
frequency-doubled Ti:Sapphire laser. This transition has a natural
linewidth of 20~MHz and is not closed since the ion may decay to
the metastable D$_{3/2}$ level. A diode laser at 866~nm serves to
repump the ion via the P$_{1/2}$ state thus closing the cooling
cycle. As the upper internal level for quantum state engineering
and sub-Doppler cooling, we employ the metastable D$_{5/2}$ level
with a natural lifetime of approx. 1~s. The
S$_{1/2}\leftrightarrow$ D$_{5/2}$ quadrupole transition at 729~nm
is excited with a Ti:Sapphire laser. We can detect whether a
transition to D$_{5/2}$ occurred by applying the beams at 397~nm
and 866~nm and monitoring the fluorescence of the ion (electron
shelving technique). The internal state of the ion is
discriminated with an efficiency close to 100$\%$ \cite{shelving}.
Another diode laser at 854 nm is used to repump the ion from the
D$_{5/2}$ level to the electronic ground state via the P$_{3/2}$
level. We observe the ions' fluorescence on a photomultiplier and
an intensified CCD camera.


\section{Spectroscopy}
For coherent spectroscopic  investigation and state engineering on
the narrow S$_{1/2} \leftrightarrow$ D$_{5/2}$ transition at 729
nm we use a pulsed technique which consists of five consecutive
steps.

\begin{enumerate}

\item{Laser light at 397~nm, 866~nm, and 854~nm is
used to pump the ion to the S$_{1/2}$ ground state. At the Doppler
limit, $E=\hbar\Gamma/2$ \cite{Wineland79}, the thermal
vibrational state corresponds to a mean vibrational quantum number
$\langle n \rangle \cong$ 10 for $\omega=(2\pi)\:1$~MHz.}
\item{A constant magnetic field of a few Gauss splits the 10 Zeeman
components of the S$_{1/2} \leftrightarrow$ D$_{5/2}$ transition
in frequency space. The S$_{1/2}(m=-1/2)$ sub-state is prepared by
optical pumping with $\sigma^-$ radiation at 397~nm. }
\item{Optional sideband cooling step (see also section 4):
The S$_{1/2}(m=-1/2) \leftrightarrow$ D$_{5/2}(m=-5/2)$ transition
is excited on one of the red sidebands. Therefore, the laser
frequency is detuned red by one motional frequency
($\omega_{laser}= \omega_{S-D} - \omega_{trap}$). The laser power
is chosen so that approximately 1~mW laser power is focused to a
waist size of 30~$\mu$m (spherical trap), or 6~$\mu$m (linear
trap). The laser at 854~nm is switched on to broaden the D$_{5/2}$
level at a power level which is set for optimum cooling. Optical
pumping to the S$_{1/2}(m=+1/2)$ level is prevented by
interspersing short laser pulses of $\sigma^-$-polarized light at
397 nm. The duration of those pulses is kept at a minimum to
prevent unwanted heating.}
\item{Spectroscopy, or alternatively, state engineering step:
We excite the S$_{1/2}(m=-1/2) \leftrightarrow$ D$_{5/2}(m=-5/2)$
transition at 729~nm with a single laser pulse, or a number of
pulses of well controlled frequency, power, and timing. These
parameters are chosen according to the desired state
manipulation.}
\item{Final state analysis: The ion's fluorescence is collected
under excitation with laser light at 397~nm and 866~nm and thus we
detect whether a transition to the shelving level D$_{5/2}$ has
been previously induced.}
\end{enumerate}

This sequence is repeated typically  100~times to measure the
D$_{5/2}$ state population $P_D$ after the state engineering step.
We study the dependence of $P_D$ on the experimental parameters,
such as the detuning $\delta\omega$ of the light at 729~nm with
respect to the ionic transition, or the length of one of the
excitation pulses in step four. The duration of a single sequence
is typically 20~ms, so we can synchronize the sequence with the ac
line frequency at 50~Hz to reduce ac-magnetic field fluctuations.

A detailed study of the Zeeman and  vibrational structure of the
S$_{1/2} \leftrightarrow$ D$_{5/2}$  transition for a single ion
in the linear trap was performed by N\"agerl et. al.
\cite{NaegerlA}. For the carrier transition (electronic excitation
only, $n \rightarrow n $) we measure a 1~kHz linewidth. We
attribute this residual broadening to laser and magnetic field
fluctuations.

\begin{figure}
\begin{minipage}{0.69\linewidth}
\begin{center}
\epsfxsize=0.98 \linewidth
\epsfbox{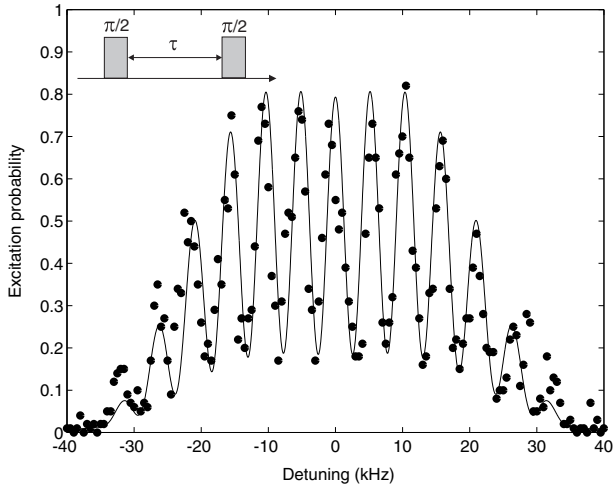}
\end{center}
\end{minipage}
\begin{minipage}{0.3\linewidth}
\caption{Ramsey spectroscopy  of a single ion on the S$_{1/2}
(m=-1/2)$ $\leftrightarrow$ D$_{5/2} (m =-5/2)$ carrier
transition. The time between  pulses was 0.2~ms and each pulse
length was 22~$\mu$s. From the fit, we deduce that the excitation
pulse length was too long for a $\pi$/2 pulse by $ \sim 10\%$ and
the purely transversal decay constant is $\sim$ 2~kHz.}
\end{minipage}
\end{figure}

Recently, we excited  Ramsey fringes using two consecutive pulses.
See Fig.~2 for the Ramsey signal of the S$_{1/2} \leftrightarrow$
D$_{5/2}$ carrier  transition for a single ion in the spherical
Paul trap. Ramsey spectroscopy allows us to investigate the
coherence of superposition states $\frac{1}{\sqrt{2}}\{
|S_{1/2}\rangle + |D_{5/2}\rangle \}$ which are excited with the
first pulse. The fit to the data of Fig. 2, allows us to estimate
the purely transversal decay constant to be 2~kHz. We attribute
this value to laser phase fluctuations and mechanical vibrations
of the trap electrodes.

In the last part of this section we will focus on spectroscopy of
ion crystals: Ions in a string are strongly coupled by the Coulomb
interaction. Small displacements from their equilibrium positions
are described in terms of normal modes of the {\em entire ion
crystal} vibrating at distinct frequencies \cite{JAMES98}. As an
example, consider two ions confined in a Paul trap. The first
normal mode corresponds to an oscillation of the entire crystal of
ions moving back and forth as if they were rigidly joined.  This
oscillation is referred to as the {\em center-of-mass mode} (COM)
\cite{JAMES98}. The second normal mode, the so-called {\em
breathing mode}, corresponds to an oscillation where the ions move
in opposite directions. The frequency of this mode is at $\sqrt{3}
\omega_{COM}$.

In the experiment {\em two ions} were trapped in the spherical
trap, and the excitation spectrum exhibits additional sideband
resonances. Apart from the center-of-mass vibration frequencies,
which remain at $\omega_x$, $\omega_y$, $\omega_z$, one observes
additional sidebands which are identified in Fig. 3. The
understanding of the full motional sideband structure presents the
basis of further experiments, such as ground state cooling and
driving coherent dynamics.

\begin{figure}
\epsfxsize=\linewidth
\epsfbox{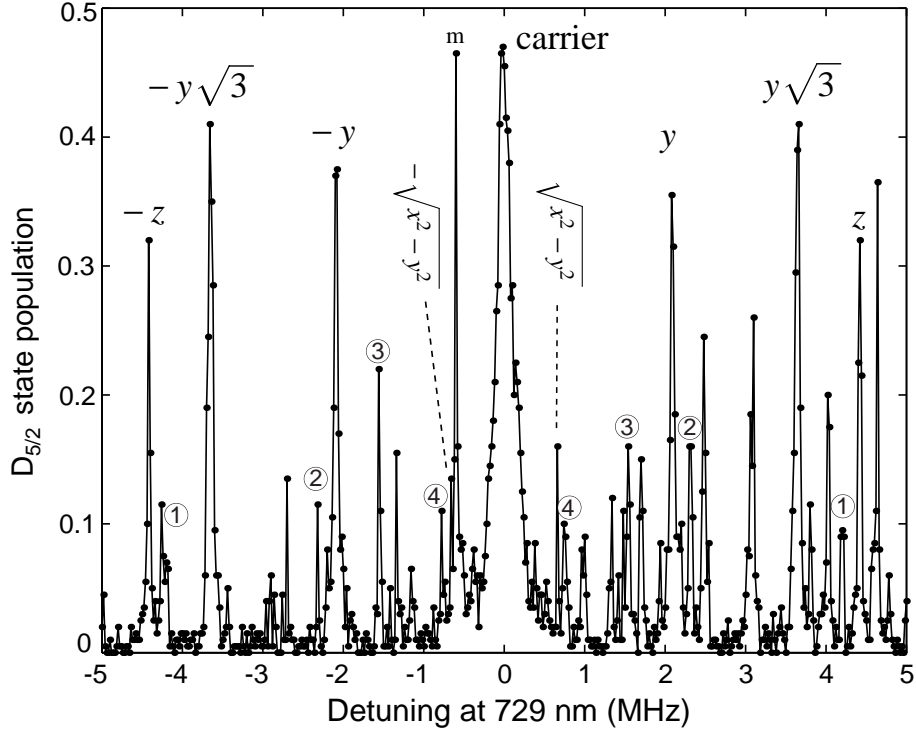}
\caption{ Sideband
excitation spectrum on the S$_{1/2} (m=-1/2)$ $\leftrightarrow$
D$_{5/2} (m =-5/2)$ for {\em two ions} in the spherical trap. All
motional sidebands are well understood in terms of normal modes:
The $\omega_y, \omega_z, \sqrt{3}\omega_y$ and $\sqrt{\omega_z^2 -
\omega_y^2}$ sidebands are clearly visible. Numbered circles
denote the second order sidebands, in detail (1): 2 $\omega_y,$
(2): $\omega_y - \omega_z$, (3): $\sqrt{3}\omega_y - \omega_z$,
(4): $\sqrt{3}\omega_z - \omega_y$. The x-direction is nearly at
right angles to the k vector of the laser beam direction and thus
only weakly excited. The line indicated with (m) belongs to a
different Zeeman transition.}
\end{figure}

\section{Cooling to the vibrational ground state}
Preparation of the motional  ground state is accomplished by a
two-stage cooling process. First, the ion is pre-cooled on the
S$_{1/2}$ - P$_{1/2}$ dipole transition. In the second stage, a
resolved-sideband cooling scheme, similar to the one used in
Ref.~\cite{Diedrich}, is applied on the narrow S$_{1/2}$ -
D$_{5/2}$ quadrupole transition: The laser frequency is detuned
from the line center by the trap frequency, $\omega_{laser} =
\omega_{SD} - \omega_{trap}$ (first "red sideband" excitation),
thus removing one phonon with each electronic excitation process.
The cooling cycle is closed by a spontaneous decay to the ground
state which conserves the phonon number in the Lamb-Dicke regime.
When the vibrational ground state $|n=0\rangle$ is reached, the
ion decouples from the laser excitation. Due to the weak coupling
between light and atom on a quadrupole transition one would expect
long cooling times. However, the cooling rate is greatly enhanced
by (i) strongly saturating the transition and (ii) shortening the
lifetime of the excited state via coupling to a dipole-allowed
transition.

\begin{figure}
\begin{center}
\epsfxsize=\linewidth
\epsfbox{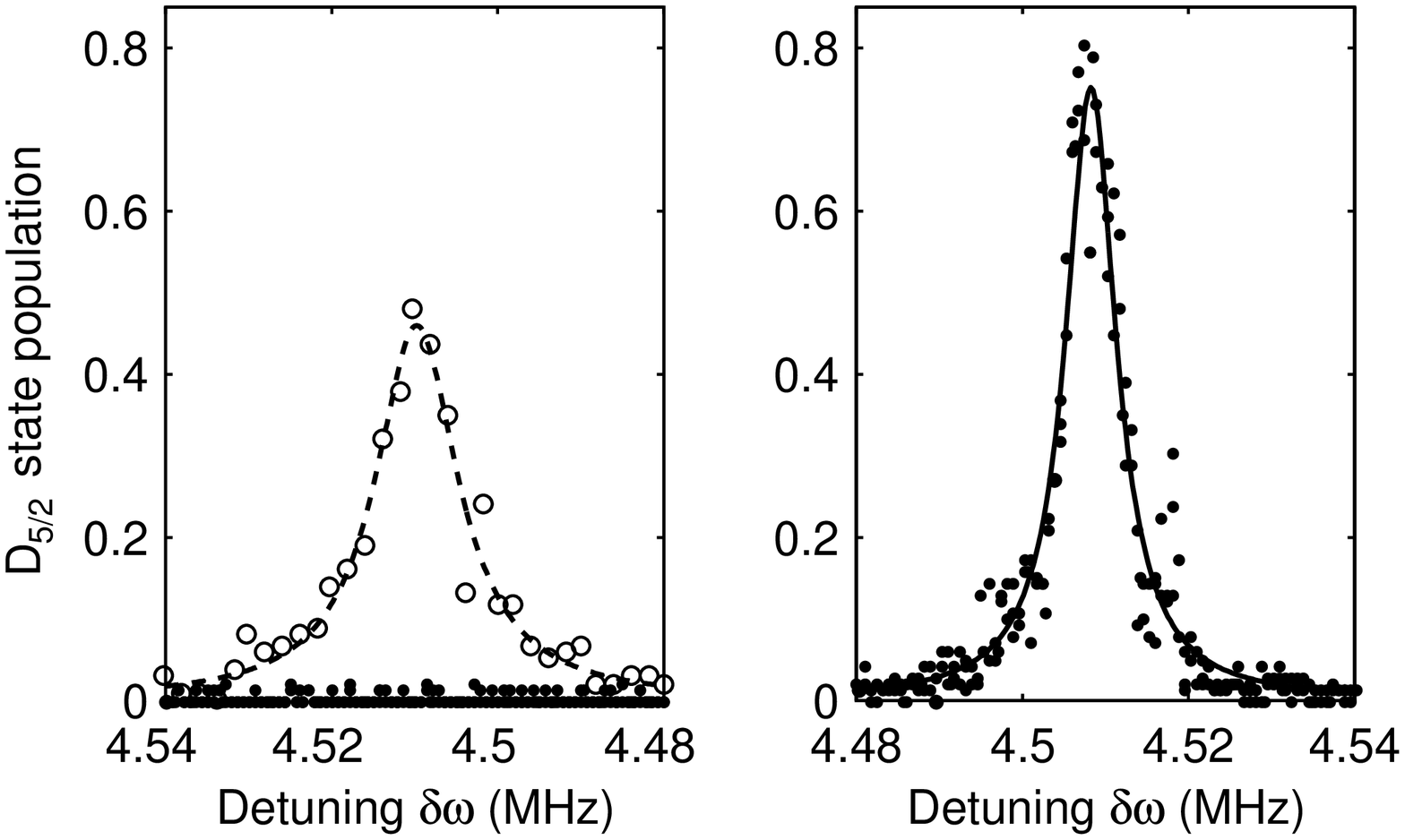}
\caption{Sideband
cooling of a single ion in the spherical Paul trap:  Sideband
absorption spectrum on the S$_{1/2} (m=-1/2)$ $\leftrightarrow$
D$_{5/2} (m =-5/2)$ transition after sideband cooling (full
circles). The frequency is centered around  the (left) red and
(right) blue sidebands at $\omega_z=4.51$~MHz. Open circles in the
left hand diagram show the red sideband after Doppler cooling
only. Each data point represents 400 individual measurements. The
analyis yields a 99.9\% ground state population.}
\end{center}
\end{figure}

After sideband cooling, the ground state occupation  is determined
by probing the sideband absorption immediately after the cooling
pulse. Fig.~4 shows the excitation probability $P_D$ for
frequencies $\omega$ centered around the red and blue $\omega_z$
sidebands. For the quantitative determination of the vibrational
ground state occupation probability $p(n=0)$ after sideband
cooling, we compare $P_D$ for excitation at $\delta\omega=-\omega$
and $\delta\omega=+\omega$, i.e. on the red and on the blue
sideband. For a thermal phonon probability distribution $p(n)$,
the ratio of excitation on the red and blue sidebands is given by
$P_ {red}/P_ {blue} = \langle n \rangle / (1+ \langle n \rangle)$.
A Lorentzian fit of the sideband heights yields a ground state
occupation of $p_0$ = 99.9$\%$ for the axial mode when $\omega_z =
(2\pi)\:4.51$~MHz \cite{Roos99}. By cooling the radial mode with
$\omega_y = (2\pi)\:2$~MHz, we transfer 98$\%$ of the population
to the motional ground state. Ground state cooling is also
possible at lower trap frequencies, however slightly less
efficient. At trap frequencies of $\omega_z = (2\pi)\:2$~MHz,
$\omega_y = (2\pi)\:0.92$~MHz we achieve $\langle n_z\rangle=0.95$
and $\langle n_y\rangle =0.85$, respectively. The x-direction is
left uncooled because it is nearly perpendicular to the cooling
beam. We also succeeded in simultaneously cooling all three
vibrational modes by using a second cooling beam and alternating
the tuning of the cooling beams between the different red
sidebands repeatedly \cite{Roos99}. For {\em two ions} in the
spherical trap we achieved ground state cooling on the
center-of-mass mode in the y and z directions, and on the
breathing-mode with more than 96$\%$ ground state probability.

In the linear trap, our goal is to implement small scale quantum
processing, thus, ions have to be kept apart so that they can be
individually addressed. Thus, the laser beam at 729~nm is tightly
focused into the linear Paul trap. We have demonstrated individual
manipulation of single ions in strings \cite{addressing}. However,
the addressing resolution is restricted by the quality of the
optical system to 6~$\mu$m. If we demand individual addressing,
this resolution restricts our ion - ion distances to greater than
6~$\mu$m, which corresponds to axial trap frequencies below
700~kHz (with up to three ions). This low trap frequency
aggravates the difficulty of ground state cooling, since the
corresponding Doppler cooling limit is close to 20~phonons. Still,
we succeed in cooling the axial mode at 700~kHz and reach $p_0$=
90$\%$ ground state. We also achieved 95$\%$ in the ground state
for cooling on the "rocking mode" at $\omega_{rock} =
\sqrt{\omega_{radial}^2 - \omega_{axial}^2}$, see Fig.~5. More
advanced techniques, such as pre-cooling on the {\em second}
sideband, prior to the usual first sideband cooling, will probably
further increase the ground state occupation.

\begin{figure}
\begin{center}
\epsfxsize=0.98 \linewidth
\epsfbox{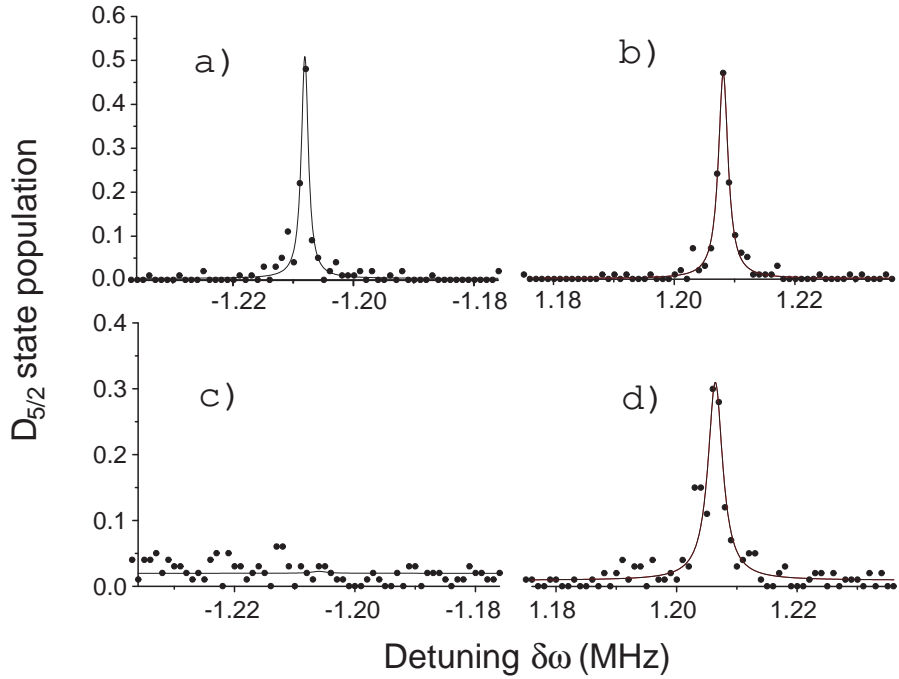}
\caption{Sideband cooling of {\em two ions} in the linear Paul
trap. Here, the ground state cooling of the "rocking" vibration
mode is shown. The frequency of the rocking mode is calculated as
$\omega_{rock} = \sqrt{\omega_{radial}^2 - \omega_{axial}^2}$
=1.212~MHz, in good agreement with the experimental finding of
1.208~MHz. We plot the excitation probability to the $D_{5/2}$
state near the sideband versus the frequency detuning of the laser
at 729~nm. In the four plots, the frequency is centered around the
red (a, c) and the blue (b, d) sideband of the rocking mode. Data
(a, b) show the sideband excitation without, (c, d) with sideband
cooling. Each data point represents 100 individual measurements.
Lorentzians are fitted to the data and we obtain a ratio
$P_{red}/P_{blue}$ which results in $p_0$ = 95(4)\% ground state
population. We achieve cooling on all radial and axial vibration
modes for two ions, and measure heating rates of $\approx$
20-50~ms per phonon.}
\end{center}
\end{figure}


\section{Coherent manipulation}

Starting from the vibrational  ground state, arbitrary quantum
states can be created. For a demonstration of  coherent state
engineering, and in order to investigate decoherence, in step 4 we
excite Rabi oscillations with the ion initially prepared in Fock
states of its motion. Radiation at 729~nm is applied on the blue
sideband transition $|$S$,n_z\rangle \leftrightarrow\:
|$D$,n_z+1\rangle$ for a given interaction time $\tau$ and the
excitation probability $P_D$ is measured as a function of $\tau$
\cite{Roos99}. The Rabi flopping behaviour allows us to analyse
the purity of the initial state and its decoherence
\cite{HarocheMeekhof}. Fig.~5a shows $P_D(\tau)$ for the
$|n=0\rangle$ state prepared by sideband cooling. Rabi
oscillations at $\Omega_{01}=(2\pi)\:21 $~kHz are observed with
high contrast indicating that coherence is maintained for times
above 1 ms.  For the preparation of the Fock state $|n=1\rangle$,
we start from $|$S$,n=0\rangle$, apply a $\pi$-pulse on the blue
sideband and an optical pumping pulse at 854 nm to transfer the
population from $|$D$,n=1\rangle$ to $|$S$, n=1\rangle$. As shown
in Fig.~5b for the $|n=1\rangle$ initial state, we also observe
Rabi oscillations with a high contrast, now at
$\Omega_{12}=(2\pi)\:30$~kHz. The Rabi frequencies have been
theoretically investigated and Blockley et. al. found
$\Omega_{n,n+1} \propto\sqrt{n+1}$ in the Lamb-Dicke regime
\cite{Blockley}, which is fulfilled in good approximation for our
trap. For the ratio of Rabi frequencies $\Omega_{12}/\Omega_{01}$
in the case of Fock states $|n=0\rangle$ and $|n=1\rangle$ we thus
expect $\sqrt{2}$. The experimental finding agrees with $\sqrt{2}$
within 1$\%$. The Fourier transform of the flopping signals also
yields directly the occupation probabilities for the contributing
Fock states $|n=0,1,2,3...\rangle$ \cite{HarocheMeekhof} and
allows us to calculate the purity of the prepared and manipulated
states. For the 'vacuum' state $|n=0\rangle$, we obtain $ p_0
=0.89(1)$ with impurites of $ p_1 = 0.09(1)$ and $ p_{n\ge 2} \le
0.02(1)$. For the Fock state $|n=1\rangle$ the populations are $
p_0 =0.03(1)$, $ p_1 =0.87(1)$, $ p_2 =0.08(2)$, and $ p_{n \ge 3}
\le 0.02(1)$. The measured transfer fidelity is about 0.9. Note
that the Rabi flopping data here were taken with less efficient
cooling (at a lower trap frequency), and the number state
occupation from the Fourier analysis is consistent with the
occupation which we determined by sideband measurements.

\begin{figure}
\epsfxsize=\linewidth
\epsfbox{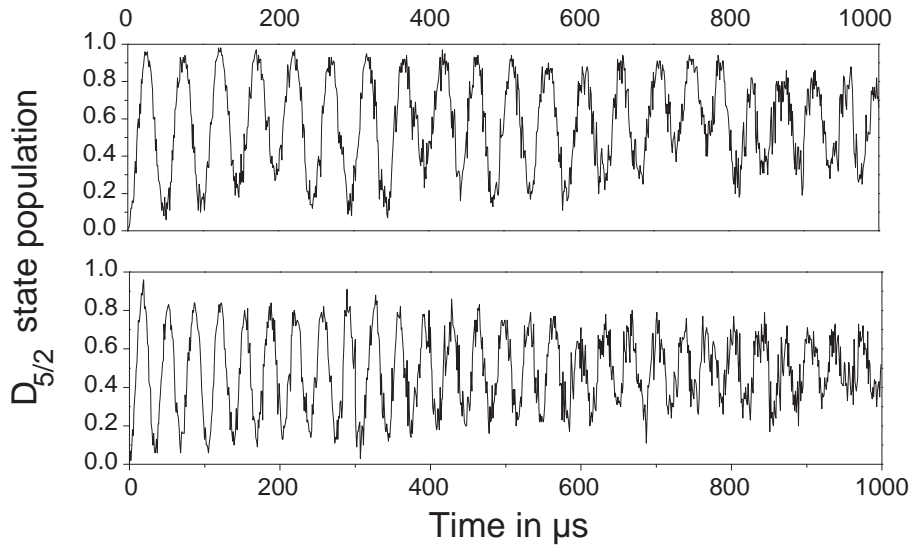}
\caption{ upper:
Rabi oscillations of a single ion in the 3D trap on the blue
sideband for the initial state $|n=0\rangle$. Coherence is
maintained for up to 1~ms. lower: Rabi oscillations as in the
upper trace, but for an initial vibrational Fock state
$|n=1\rangle$. }
\end{figure}

\section{Speed limits of gate operations}
As a figure of merit for an ion trap quantum processor, we
calculate how many gate operations are feasible within the time
scale given by decoherence. We have investigated the speed limit
of quantum gate operations. The proposed methods
\cite{CiracZoller,MonroeGate} employ the use of laser pulses to
entangle the electronic and the vibrational degrees of freedom of
trapped ions. A theoretical investigation shows that the proposed
methods are limited mainly by the recoil frequency of the relevant
electronic transition \cite{fastgate}. We have experimentally
studied the basic building block of a gate operation, that is a
$\pi$-pulse excitation on the first blue sideband. If the blue
sideband is driven significantly faster than the inverse of the
recoil frequency, carrier transitions are excited off-resonance
and the contrast of the observed Rabi oscillations decreases. In
the case of the quadrupole S$_{1/2} \leftrightarrow$ D$_{5/2}$
transition we find a time of about 16~$\mu$s for this limit, if we
demand a gate fidelity of 99~$\%$. For further details on fast
gates, we refer to A. Steane et. al. \cite{fastgate}. We thus
estimate that 30 - 50 quantum gate operations will be feasible
with the current setup.


\section{Conclusion}
We have  engineered the quantum states of motion $|n=0,1 \rangle$
of a single trapped ion that are relevant for quantum computation,
using laser excitation on a forbidden optical transition. We have
observed more than 30 periods of Rabi oscillations on the motional
sidebands of this transition, thus showing that decoherence is
negligible on the time scale of a few oscillations, i.e. the time
required for a quantum gate operation. We attribute the observed
1~ms decoherence time to laser and magnetic field fluctuations.
Heating of the motional degrees of freedom has also been observed
and was measured directly to happen at least 1 order of magnitude
more slowly. This confirms that in the comparatively large traps
which we use, heating seems not to be a limiting process.

Furthermore, we have  been  able to cool two ions to the ground
state, and also to reach the ground state for single and two ions
in the linear trap, under conditions which allow individual
addressing of ions in a string. Multiple coherent gate operations
with trapped ions seem well within experimental reach.

This work is supported by the Austrian  "Fonds zur F\"orderung der
wissenschaftlichen Forschung" within the project SFB15, and in
parts by the European Commission within the TMR networks "Quantum
Information" (ERB-FMRX-CT96-0087) and "Quantum Structures"
(ERB-FMRX-CT96-0077), and the "Institut f\"ur Quanteninformation
GmbH".

\end{document}